\documentclass[a4paper,11pt]{article}
\usepackage{pos,subcaption}
\usepackage{setspace}
\title{The Fluorescence Camera for the PBR mission}
\ShortTitle{The FC for PBR}

\manuallySeparateAuthors

\author*[a]{Francesco S. Cafagna}
\onbehalf{for the JEM-EUSO Collaboration\\[-1mm]{\normalsize \normalfont (a complete list of authors can be found at the end of the proceedings)}}

\affiliation[a]{Istituto Nazionale di Fisica Nucleare -  Sezione di Bari,\\
  Via Orabona 4, Bari, Italy}

\emailAdd{francesco.cafagna@ba.infn.it}

\abstract{The Probe Of Extreme Multi-Messenger Astrophysics (POEMMA) Balloon with Radio (PBR) is an instrument designed to be borne by a NASA suborbital Super Pressure Balloon (SPB), in a mission planned to last as long as 50 days. The PBR instrument consists of a 1.1 m aperture Schmidt telescope, similar to the POEMMA design, with two cameras in its hybrid focal surface: a Fluorescence Camera (FC) and a Cherenkov Camera (CC), both mounted on a frame that can be tilted to point from nadir up to 13 degrees above the horizon. The FC camera is designed to detect the fluorescence emission of Extensive Air Showers produced by Ultra-High Energy Cosmic Rays from sub-orbital altitudes. This measurement will validate the detection strategy for future space-based missions, such as POEMMA. The FC will be made of 4 Photo Detection Modules (PDMs), each consisting of a 6×6 matrix of  64-channel Multi Anode PhotoMulTipliers (MAPMT), for a grand total of 2304 pixels for each PDM. Custom-designed SPACIROC-3 ASICs perform single photoelectron counting on each pixel as well as charge integration on groups of 8 pixels to measure extremely bright or fast signals, reaching a double pulse resolution in the order of 10 ns for a 1 microsecond acquisition gate. A field flattener lens and a BG3 filter, to match the wavelength range of interest (300-400 nm), are mounted in front of the PDM. The camera will be able to detect showers in a field of view of 24x24 square degrees, with a pixel size on ground corresponding to 115 m. Details on the camera design and implementation will be given, along with the expected performance and the state of the construction.}

\ConferenceLogo{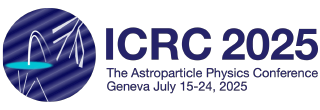}

\FullConference{39th International Cosmic Ray Conference (ICRC2025)\\
 15–24 July 2025\\
Geneva, Switzerland\\}

\begin{document}
\maketitle

\section{Introduction}
\label{sec:intro}

The study of the origin and accelerating mechanisms of Ultra-High-Energy Cosmic Rays (UHECRs) is still an open question that should be addressed~\cite{Coleman2023} by the community.
These scientific goals can be obtained by using high-accuracy detectors and by maximizing the exposure at the highest energies.
  
To address these questions, the JEM-EUSO (Joint Exploratory Missions for Extreme Universe
Space Observatory) international Collaboration\footnote{\url{https://www.jemeuso.org}} has developed a program to observe UHECR ($E>10$EeV) and High-Energy (HE) neutrino ($E>1$PeV) from space~\cite{Plebaniak_JEMEUSO_icrc2025}. 
The Collaboration's main objective is to
develop a large mission with dedicated instrumentation looking down on the Earth’s 
atmosphere from space, both towards the nadir and/or
towards the limb, to detect the Extensive Air 
Showers (EAS) initiated by the above-mentioned particles in the higher atmosphere. 

The JEM-EUSO approach aims to increase the exposure, thanks to the observation from space, to UHECRs and to achieve a near-uniform full sky coverage. 
Moreover the instrument design is optimized to obtain, with a single detector, an unprecedented statistical accuracy minimizing systematic uncertainties.

Besides these major scientific goals, the indirect measurements from space could also be used to study atmospheric and space-weather sciences (like the characterization of night-glow and transient luminous events, meteors and
meteoroids or the tracking of space debris), as secondary scientific goals~\cite{Caruso:2025Ms}.

The JEM-EUSO collaboration proposed a dual satellite mission:
Probe for Multi-Messenger Astrophysics (POEMMA)~\cite{Olinto2021},  
which proposes the use of a hybrid focal surface to detect UV and Cherenkov light from space.

Due to the complexity of the chosen detectors and mission itself, the actual Collaboration 
efforts are aimed at designing and building pathfinder
missions that do improve the technological readiness and validate the detection techniques. 
A long duration stratospheric balloon offers an opportunity for such a mission in a near-space
environment without the risk and cost of a satellite space mission. 

The POEMMA-Balloon with Radio (PBR)~\cite{Eser:2025/T,Eser_PBR_icrc2025} payload is such a pathfinder mission, based on the design study of the POEMMA and previous long duration balloon missions, namely: EUSO-SPB1 \cite{Abdellaoui2024} and EUSO-SPB2~\cite{Eser:2023lck}, 
and Mini-EUSO
~\cite{Bacholle:2020emk}, a pathfinder borne in the International Space Station (ISS). 

\section{The POEMMA-Balloon with Radio payload}

The PBR payload is designed to fly on a NASA Super Pressure balloon and to sustain a flight as long as 50 days. 
The mission is planned to launch from Wanaka, NZ in the first half of 2027 and will circulate the Southern Ocean. 

\begin{figure}
	\begin{subfigure}[c]{0.5\columnwidth}
		\centering
		\includegraphics[width=.95\columnwidth]{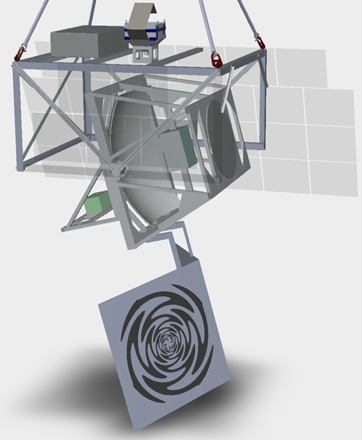}
		\caption{The PBR payload}
		\label{fig:pbr}
	\end{subfigure}\hfill
	\begin{subfigure}[c]{0.5\columnwidth}
		\centering
		\includegraphics[width=.95\columnwidth]{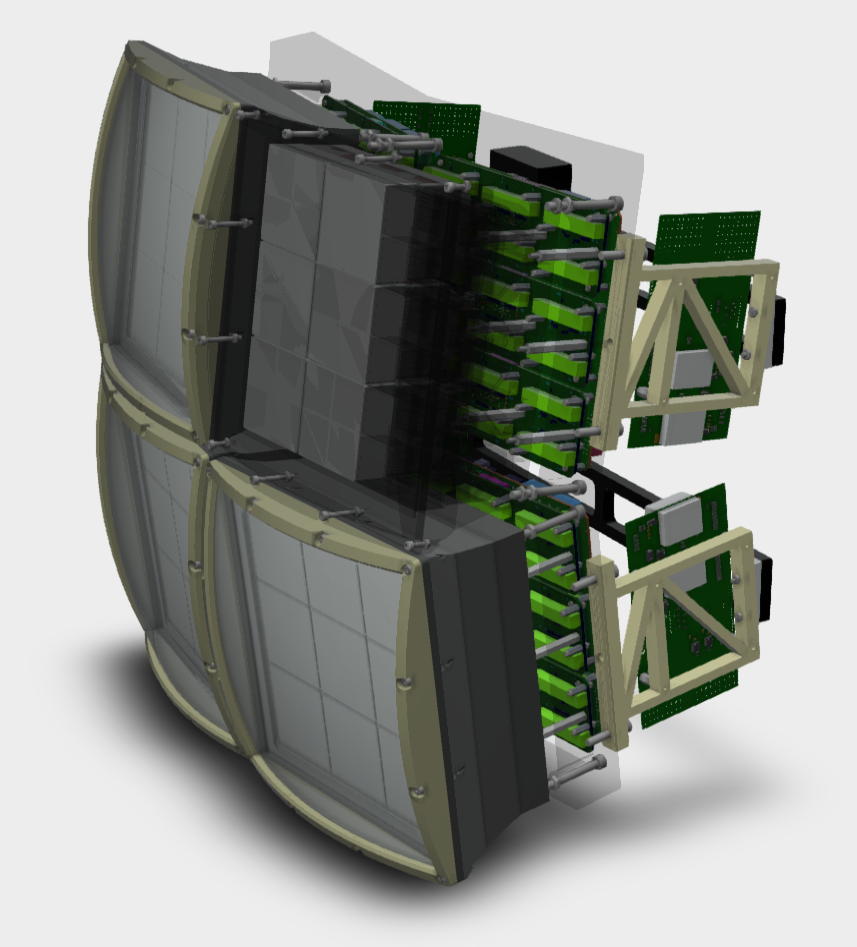}
		\caption{The PBR Fluorescence Camera}
		\label{fig:fc} 
	\end{subfigure}
	\caption{ (\subref{fig:pbr}) The PBR payload without the containment boxes, ancillary detectors and electronic boxes. (\subref{fig:fc}) The drawing of the FC camera. Part of the mechanical frame is shaded along with some component of the top right Photo Detection Module (PDM).}
	\label{fig:pbr_fc}
\end{figure}

The payload consists of a tiltable telescope, housing a hybrid focal surface, along with a Radio Instrument (RI) made of two antennas mounted beneath the
telescope. A drawing of the bare structure of the payload and its various components is shown in Fig.\ref{fig:pbr}.

The PBR hybrid focal surface is composed of a 
Fluorescence Camera (FC), along with a Cherenkov 
Camera (CC)~\cite{Scotti_CC_icrc2025}. The sensitive surface is seated on the focus of a 1.1m aperture Schmidt telescope. An aspheric correction plate, made from polymethyl methacrylate (PMMA), closes the pupil 
that focuses photons on the focal surface by means
of a mirror, measuring roughly 2$\times$2m in size, composed by 12 segments mounted in a 3 rows by 4 columns configuration. The mirror has a $\sim$1.6m radius of curvature and will provide a field of view (FoV) of $\sim$36${^\circ}\times$30${^\circ}$, with an expected point spread function (PSF) of 3mm in diameter (at 95\% containment).
The reflective surface is vacuum-deposited aluminum with a film thickness chosen to optimize reflectivity in the UV, which is protected by a silicon dioxide coating.
The FC and CC electronics, will be attached to the tiltable telescope frame that will be closed in a light-tight box (not shown in the Fig.~\ref{fig:pbr})~\cite{Mayotte_MECH_icrc2025}. A remote controlled tilt system will actuate the box tilt up to an angle of 13${^\circ}$ above the horizon.

The Radio Instrument (RI) antennas will be attached to the bottom of the telescope box. The RI design is inherited from the PUEO LF~\cite{Abarr2021} instrument and comprises two radio receivers, each featuring dual-polarized sinuous antennas, a two-stage front-end RF signal conditioning chain, and associated power systems.

In addition to these main instruments, PBR will be equipped with an infrared camera to monitor clouds in the instruments FoV and will, for the first time, include a $\gamma$ and $X$-ray detector, opening yet another detection channel to investigate a wide range of phenomena.

\label{sec:pbr}

\section{The Fluorescence Camera}
\label{sec:fc}

The PBR FC camera is designed to detect the fluorescence emission of Extensive Air Showers (EAS) produced by UHECR from sub-orbital altitudes. The FC design is based on the technology developed by the JEM-EUSO collaboration over the last decade and used in all the pathfinders constructed until now. 

The FC is modular in design and consists of 4 Photo Detection Modules (PDMs). 
The PDM is the core module of the JEM-EUSO
fluorescence detection technique.
The PDM used in PBR is the evolution of the PDM designed and built for every pathfinder constructed by the Collaboration~\cite{ADAMS2025103046}. The PDMs will be arranged in a matrix of 2$\times$2 as shown in Fig~\ref{fig:ec}. The PDM matrix will be fixed to an aluminum frame that will interface the camera to the telescope mechanical frame. The telescope mechanical frame will house the FC Data Processor (DP) that comprises data acquisition (DAQ) CPUs and electronics, along with housekeeping boards and several services. 

\subsection{The Photo Detection Module} 

The PDM is a standalone UV camera that counts photoelectrons in a 1.05$\mu$s gate. Its sensitive plane is made of 2304 pixels, arranged in a 48$\times$48 pixel matrix.
The pixel matrix consists of a 6$\times$6 matrix of 64-channel Multi Anode PhotoMulTipliers (MAPMTs). 
MAPMTs are grouped in groups of 4 in a so-called Elementary Cell (EC).
An EC, besides the 4 MAPMTs, houses the front-end electronics and the high voltage distribution board. A grand total of 9 ECs, arranged in a 3$\times$3 matrix, are contained in a PDM. Each EC row is connected to a board that implements the multiplexing of the front-end electronics data and control links with the read-out board. This board implements the first level trigger logic and the read-out of the whole PDM along with the control and monitoring of the PDM
high voltage supply. A dedicated board will supply the High Voltages for each PDM; such a board is based on a Cockcroft–Walton circuit delivering up to 1100V.

Each first level trigger is collected by a trigger and synchronization board that broadcast the global first level trigger to all the read-out boards to start data acquisition. This board generates the global gate time unit (GTU) signal that is used by the front-end electronics and does estimate the live and dead time of the FC. 
Data generated upon global trigger request, is transferred to a CPU that merges the PDMs data into a single event record. 
The CPU also implements the overall monitor and control, receiving the commands to operate the FC and broadcasting the housekeeping data. Housekeeping data includes not only data collected by a dedicated board but also the status of every FC component along with the log of the CPU's DAQ processes.  

A 3D printed mechanical frame houses the 9 ECs,
the multiplexing and read-out boards and the high voltage supply. The frame is made out of a special honeycomb structure, to reduce the weight while complying to the requested mechanical requirements. The material chosen is a micro carbon fiber filled nylon: the ONYX${^\circledR}$\footnote{\url{https://markforged.com/materials/plastics/onyx} }, inserting all the needed small parts during the print process. 

A field flattener lens and a BG3 filter, to match the wavelength range of interest (300-400 nm), will be mounted in front of a PDM; 
they will be fixed on the mechanical frame. The FC will be able to detect showers in a field of view of 24°x24°, with a pixel size on ground corresponding to $\sim$115m. 

\begin{figure}
	\centering
	\includegraphics[width=.5\columnwidth]{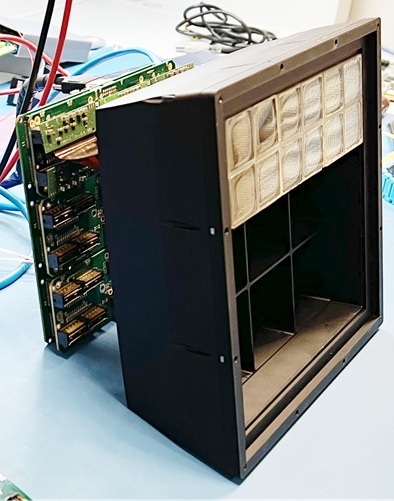}
	\caption{
    A PDM prototype under test.
    Only three Elementary Cells (ECs) are mounted.
    The read-out electronics is attached on the back of the 3D printed mechanical frame.}
	\label{fig:pdm_test}
\end{figure}

The first prototype of a PDM, during a first integration test, is shown in Fig.~\ref{fig:pdm_test}. 
The frame is filled with just three ECs out of nine, while the multiplexer board and the read-out board are attached to the back;
no lens and filter are shown in the picture. 

\subsection{The Elementary Cell}

\begin{figure}
	\begin{subfigure}[c]{0.5\columnwidth}
		\centering
		\includegraphics[width=1\columnwidth]{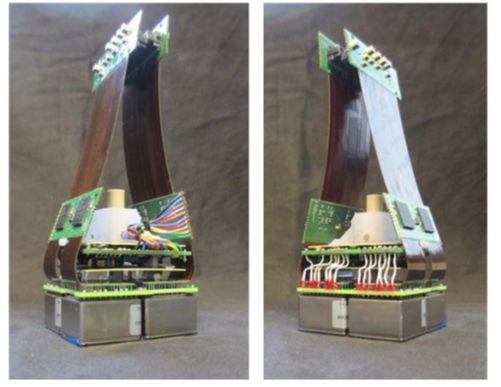}
		\caption{
        The EC without potting.
        }
		\label{fig:ec_nopot}    
	\end{subfigure}\hfill
	\begin{subfigure}[c]{0.5\columnwidth}
		\centering
		\includegraphics[width=.8\columnwidth]{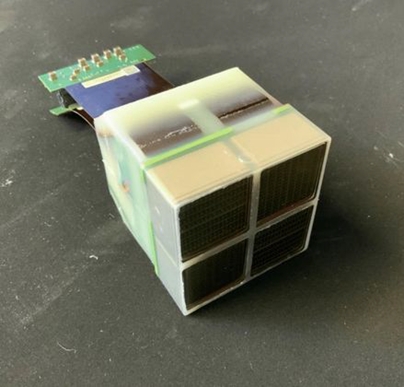}
		\caption{The EC with the potting.} 
		\label{fig:ec_potted}
	\end{subfigure}
	\caption{(\subref{fig:ec_nopot}) An Elementary Cell before potting. The ASICs are visible mounted on the lateral boards that are integrated with kapton flat cables on both board sides. (\subref{fig:ec_potted}) An Elementary Cell after the potting, ready to be integrated in the PDM frame. }
	\label{fig:ec}
\end{figure}

An Elementary Cell (EC) houses four 64-pixel MAPMTs (Hamamatsu R11265-03-M64, Ultra Bialkali).
Besides, the EC accommodates also the detectors front-end and high voltage distribution electronics. All these components are assembled in a compact format, as shown in Fig.~\ref{fig:ec}. In particular, Fig.~\ref{fig:ec_nopot} shows the inner structure of an EC before the potting. The assembled EC is potted using a space qualified epoxy compound to prevent discharge between the various components and isolate the EC itself; 
a potted EC is shown in
Fig.~\ref{fig:ec_potted}.

The front-end electronics is distributed on 4 boards for each side, 
connected to 2 MAPMTs at the same time.
The first board is soldered to the MAPMT and the high voltage distribution board. 
Two of such boards
are connected by a kapton flat cable to the front-end chip board, 
housing two front-end chips. This board is connected to a connection board, by another kapton flat cable. This cable carries the data and control links of the front-end chip, along with the high voltage lines running from the read-out board to the MAPMTs. 

The front-end chip is a custom-designed SPACIROC-3~\cite{Blin:2018tjp} ASICs able to perform 
single-photoelectron counting on each pixel as well as charge integration on groups of 8 pixels to measure extremely bright or fast signals. This chip performs a double pulse resolution of the order of 10 ns for a 1.05$\mu$s acquisition gate (GTU). 

\subsection{The read-out electronics}

\begin{figure}
	\begin{subfigure}[c]{0.5\columnwidth}
		\centering 
	    \includegraphics[width=.75\columnwidth]{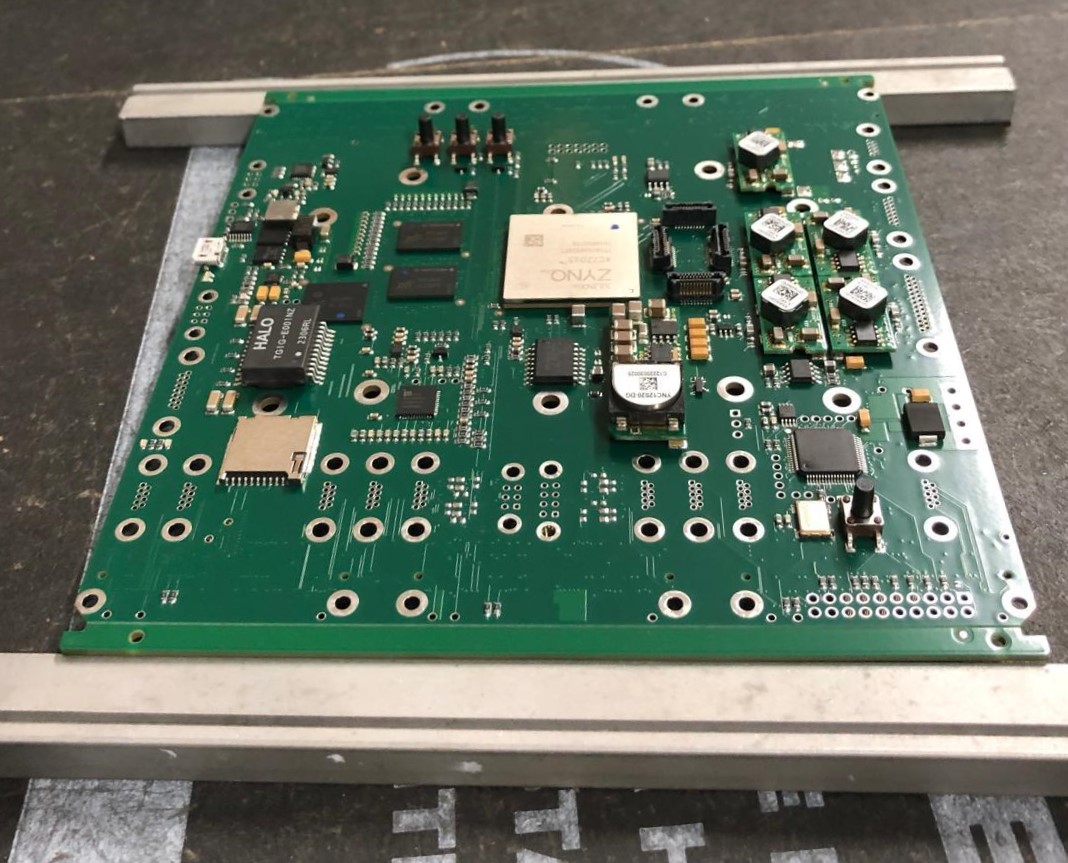}
	    \caption{The PDM read-out board}
	    \label{fig:zynq}
	\end{subfigure}\hfill
	\begin{subfigure}[c]{0.5\columnwidth}
		\centering 
		\includegraphics[width=.5\columnwidth]{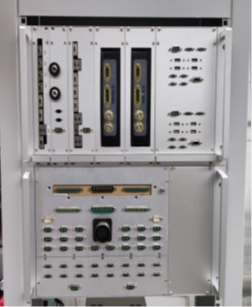}
		\caption{The FC Data Processor}
		\label{fig:dp}
	\end{subfigure}
    \caption{(\subref{fig:zynq}) The AMD/Xilinx Zynq based read-out board of a PDM. (\subref{fig:dp}) A partial view of the FC Data Processor. The second CPU is not shown in the picture.
    }
\end{figure}

A grand total of 36 front-end ASICs are present in
each PDM and read-out by a single board. The link between the ASICs of each EC and the read-out board are multiplexed by a total of three boards, 
each of these boards
is equipped with an AMD/Xilinx Artix 7 FPGA. These three boards are connected to the main
read-out board which houses an AMD/Xilinx Zynq 7000 FPGA featuring an embedded dual core ARM9 CPU processing system. 
The read-out board controls the ASIC data flows and
configurations, implements the trigger logics, for both the photon counting and charge integrated working mode, and is interfaced with the DP CPU for the data storage, PDM monitor and remote control. Moreover, it does manage the EC's high voltage distribution cards. The trigger algorithm, designed to look for clusters of signals over threshold, was developed and worked for the previous EUSO-SPB2 mission~\cite{Filippatos:2022zfl}. The first prototype of the PBR PDM read-out board is shown in Fig.~\ref{fig:zynq}; the same board, attached to the three multiplexing boards, 
can be seen at the back of the PDM frame shown in Fig.~\ref{fig:pdm_test}.

The level one PDM triggers are collected by a trigger and clock board housed in the DP. 
This board generates the global trigger and broadcasts it to all the PDMs.
Upon trigger reception, the read-out board starts the read-out of a buffer of the data collected in a 
time window large 128 GTU wide, centered on the GTU corresponding to the time of the global trigger signal.   

\subsection{The Data Processor} 

The Data Processor (DP) generates the global trigger, the busy and clock reference signals\cite{Scotti_DP_icrc2025}. 
Moreover, it controls the FC and ensures the storage
and the filter of the housekeeping and scientific data, and implements the interface with the payload telemetry. The DP houses: 2 redundant CPUs, a trigger and synchronization board, 2 differential GPS receivers, 
a housekeeping board, 4 Ethernet switches;
and 2 Solid State Power Distribution Units (PDU). 

Besides hosting the trigger and clock synchronization board, the DP houses a housekeeping board that reads out several sensors installed on each PDM and the focal surface. The housekeeping board is interfaced also with two differential GPS 
that can be read out by the CPUs.

The 2 redundant CPUs will receive and handle remote commands, from the telemetry system, and will 
implement the run control, filtering and transfer
of all the scientific and housekeeping data collected from the FC sub-systems. 

\section{Conclusion}

PBR is a pathfinder for the POEMMA observatory, equipped with a Radio detector designed to fly on a Super-Pressure Balloon.
For the first time an optical telescope and a radio detector will be used together on a balloon borne apparatus. 
The optical telescope will be equipped with a Fluorescence Camera (FC) to detect UHECR measuring the fluorescence component of the cosmic ray induced showers. 
The FC is under construction, while first tests and commissioning are ongoing to meet the goal of a flight in the spring of 2027.

\section*{Acknowledgements}

The authors would like to acknowledge the support by NASA award 80NSSC22K1488 and 80NSSC24K1780, by the French space agency CNES and the 
Italian Space Agency ASI.
The work is supported by OP JAC financed by ESIF and the MEYS CZ.02.01.01/00/22\_008/0004596. 
We gratefully acknowledge the contributions and expert advice of the PUEO collaboration.
We also acknowledge the invaluable contributions of the administrative and technical staff at our home institutions.

\bibliographystyle{JHEP}
\bibliography{pbr.bib}

@article{Scotti_DP_icrc2025,
	author = "Scotti, Valetina",
	collaboration = "JEM-EUSO",
	title = "{The Data Processor system of the PBR mission}",
	journal = "PoS",
	volume = "ICRC2025",
	pages = "527",
	year = "2025"
}

@article{Scotti_CC_icrc2025,
	author = "Scotti, Valetina",
	collaboration = "JEM-EUSO",
	title = "{The Cherenkov Camera for the PBR mission}",
	journal = "PoS",
	volume = "ICRC2025",
	pages = "391",
	year = "2025"
}

@article{Eser_PBR_icrc2025,
	author = "Eser, Johannes",
	collaboration = "JEM-EUSO",
	title = "{POEMMA-Balloon with Radio: An Overview}",
	journal = "PoS",
	volume = "ICRC2025",
	pages = "249",
	year = "2025"
}

@article{Mayotte_MECH_icrc2025,
	author = "Mayotte, Eric",
	collaboration = "JEM-EUSO",
	title = "{The Optical and Mechanical Design of POEMMA Balloon with Radio}",
	journal = "PoS",
	volume = "ICRC2025",
	pages = "332",
	year = "2025"
}

@article{Plebaniak_JEMEUSO_icrc2025,
	author = "Plebaniak, Zbigniew",
	collaboration = "JEM-EUSO",
	title = "{From Ground to Space: An Overview of the JEM-EUSO Program for the Study of UHECRs and Astrophysical Neutrinos}",
	journal = "PoS",
	volume = "ICRC2025",
	pages = "360",
	year = "2025"
}

@Article{Abdellaoui2024,
  author        = {Abdellaoui, G. and others},
  journal       = {Astroparticle Physics},
  title         = {{EUSO-SPB1 mission and science}},
  year          = {2024},
  issn          = {0927-6505},
  month         = jan,
  pages         = {102891},
  volume        = {154},
  collaboration = {JEM-EUSO},
  doi           = {10.1016/j.astropartphys.2023.102891},
  publisher     = {Elsevier BV},
  url           = {http://dx.doi.org/10.1016/j.astropartphys.2023.102891},
}

@article{Eser:2023lck,
    author = "Eser, Johannes and Olinto, Angela V. and Wiencke, Lawrence",
    collaboration = "JEM-EUSO",
    title = "{Overview and First Results of EUSO-SPB2}",
    eprint = "2308.15693",
    archivePrefix = "arXiv",
    primaryClass = "astro-ph.HE",
    doi = "10.22323/1.444.0397",
    journal = "PoS",
    volume = "ICRC2023",
    pages = "397",
    year = "2023"
}

@article{ADAMS2025103046,
	title = {The EUSO-SPB2 fluorescence telescope for the detection of Ultra-High Energy Cosmic Rays},
	journal = {Astroparticle Physics},
	volume = {165},
	pages = {103046},
	year = {2025},
	issn = {0927-6505},
	doi = {https://doi.org/10.1016/j.astropartphys.2024.103046},
	url = {https://www.sciencedirect.com/science/article/pii/S0927650524001233},
	author = {James H. Adams and Denis Allard and Phillip Alldredge and Luis Anchordoqui and Anna Anzalone and Matteo Battisti and Alexander A. Belov and Mario Bertaina and Peter F. Bertone and Sylvie Blin-Bondil and Julia Burton and Francesco S. Cafagna and Marco Casolino and Karel Černý and Mark J. Christl and Roberta Colalillo and Hank J. Crawford and Alexandre Creusot and Austin Cummings and Rebecca Diesing and Alessandro Di Nola and Toshikazu Ebisuzaki and Johannes Eser and Silvia Ferrarese and George Filippatos and William W. Finch and Flavia Flaminio and Claudio Fornaro and Duncan Fuehne and Christer Fuglesang and Diksha Garg and Alessio Golzio and Fausto Guarino and Claire Guépin and Tobias Heibges and Eleanor G. Judd and Pavel A. Klimov and John F. Krizmanic and Viktoria Kungel and Luke Kupari and Evgeny Kuznetsov and Massimiliano Manfrin and Włodzimierz Marszał and John N. Matthews and Marco Mese and Stephan S. Meyer and Marco Mignone and Hiroko Miyamoto and Alexey S. Murashov and Jane M. Nachtman and Angela V. Olinto and Yasar Onel and Giuseppe Osteria and Beatrice Panico and Ètienne Parizot and Tom Paul and Miroslav Pech and Francesco Perfetto and Lech W. Piotrowski and Zbigniew Plebaniak and Jonatan Posligua and Guillaume Prévôt and Marika Przybylak and Patrick Reardon and Mary Hall Reno and Marco Ricci and Fred Sarazin and P. Schovánek and Valentina Scotti and Kenji Shinozaki and Jorge F. Soriano and Ben K. Stillwell and Jacek Szabelski and Yoshiyuki Takizawa and Daniil Trofimov and Fredrik Unel and Laura Valore and Tonia M. Venters and John Watts and Lawrence Wiencke and Hannah Wistrand and Roy Young}
}

@article{Blin:2018tjp,
    author = "Blin, S. and Barrillon, P. and de La Taille, C. and Dulucq, F. and Gorodetzky, P. and Pr\'ev\^ot, G.",
    collaboration = "JEM-EUSO",
    title = "{SPACIROC3: 100 MHz photon counting ASIC for EUSO-SPB}",
    doi = "10.1016/j.nima.2017.12.060",
    journal = "Nucl. Instrum. Meth. A",
    volume = "912",
    pages = "363--367",
    year = "2018"
}

@article{Filippatos:2022zfl,
    author = "Filippatos, George and Battisti, Matteo and Belov, Alexander and Bertaina, Mario and Bisconti, Francesca and Eser, Johannes and Mignone, Marco and Sarazin, Fred and Wiencke, Lawrence",
    title = "{Development of a cosmic ray oriented trigger for the fluorescence telescope on EUSO-SPB2}",
    eprint = "2201.00794",
    archivePrefix = "arXiv",
    primaryClass = "astro-ph.HE",
    doi = "10.1016/j.asr.2021.12.028",
    journal = "Adv. Space Res.",
    volume = "70",
    pages = "2794--2803",
    year = "2022"
}

@Article{Olinto2021,
  author        = {Olinto, A. V. and others},
  journal       = {JCAP},
  title         = {{The POEMMA (Probe of Extreme Multi-Messenger Astrophysics) observatory}},
  year          = {2021},
  pages         = {007},
  volume        = {06},
  archiveprefix = {arXiv},
  collaboration = {POEMMA},
  doi           = {10.1088/1475-7516/2021/06/007},
  eprint        = {2012.07945},
  primaryclass  = {astro-ph.IM},
}

@Article{Abarr2021,
  author        = {Abarr, Q. and others},
  journal       = {JINST},
  title         = {{The Payload for Ultrahigh Energy Observations (PUEO): a white paper}},
  year          = {2021},
  number        = {08},
  pages         = {P08035},
  volume        = {16},
  archiveprefix = {arXiv},
  collaboration = {PUEO},
  doi           = {10.1088/1748-0221/16/08/P08035},
  eprint        = {2010.02892},
  primaryclass  = {astro-ph.IM},
}

@article{Bacholle:2020emk,
    author = "Bacholle, S. and others",
    title = "{Mini-EUSO Mission to Study Earth UV Emissions on board the ISS}",
    eprint = "2010.01937",
    archivePrefix = "arXiv",
    primaryClass = "astro-ph.IM",
    doi = "10.3847/1538-4365/abd93d",
    journal = "Astrophys. J. Suppl.",
    volume = "253",
    number = "2",
    pages = "36",
    year = "2021"
}

@Article{Eser:2025/T,
  author  = {Eser, Johannes and Mayotte, E. and Olinto, Angela V. and Osteria, Guiseppe},
  journal = {PoS},
  title   = {{POEMMA-Balloon with Radio, towards a space-based Multi-Messenger Observatory}},
  year    = {2025},
  pages   = {061},
  volume  = {UHECR2024},
  doi     = {10.22323/1.484.0061},
}

@Article{Caruso:2025Ms,
  author  = {Caruso, Rossella},
  journal = {PoS},
  title   = {{Overview of the JEM-EUSO Program}},
  year    = {2025},
  pages   = {060},
  volume  = {UHECR2024},
  doi     = {10.22323/1.484.0060},
}

@Article{Coleman2023,
  author   = {A. Coleman and J. Eser and E. Mayotte and F. Sarazin and F.G. Schröder and D. Soldin and T.M. Venters and R. Aloisio and J. Alvarez-Muñiz and R. {Alves Batista} and D. Bergman and M. Bertaina and L. Caccianiga and O. Deligny and H.P. Dembinski and P.B. Denton and A. {di Matteo} and N. Globus and J. Glombitza and G. Golup and A. Haungs and J.R. Hörandel and T.R. Jaffe and J.L. Kelley and J.F. Krizmanic and L. Lu and J.N. Matthews and I. Mariş and R. Mussa and F. Oikonomou and T. Pierog and E. Santos and P. Tinyakov and Y. Tsunesada and M. Unger and A. Yushkov and M.G. Albrow and L.A. Anchordoqui and K. Andeen and E. Arnone and D. Barghini and E. Bechtol and J.A. Bellido and M. Casolino and A. Castellina and L. Cazon and R. Conceição and R. Cremonini and H. Dujmovic and R. Engel and G. Farrar and F. Fenu and S. Ferrarese and T. Fujii and D. Gardiol and M. Gritsevich and P. Homola and T. Huege and K.-H. Kampert and D. Kang and E. Kido and P. Klimov and K. Kotera and B. Kozelov and A. Leszczyńska and J. Madsen and L. Marcelli and M. Marisaldi and O. Martineau-Huynh and S. Mayotte and K. Mulrey and K. Murase and M.S. Muzio and S. Ogio and A.V. Olinto and Y. Onel and T. Paul and L. Piotrowski and M. Plum and B. Pont and M. Reininghaus and B. Riedel and F. Riehn and M. Roth and T. Sako and F. Schlüter and D.H. Shoemaker and J. Sidhu and I. Sidelnik and C. Timmermans and O. Tkachenko and D. Veberic and S. Verpoest and V. Verzi and J. Vícha and D. Winn and E. Zas and M. Zotov},
  journal  = {Astroparticle Physics},
  title    = {Ultra high energy cosmic rays The intersection of the Cosmic and Energy Frontiers},
  year     = {2023},
  issn     = {0927-6505},
  pages    = {102819},
  volume   = {149},
  doi      = {https://doi.org/10.1016/j.astropartphys.2023.102819},
  url      = {https://www.sciencedirect.com/science/article/pii/S0927650523000051},
}

\newpage
\begin{center}
{\Large\bf Full Authors list: The JEM-EUSO Collaboration}	
\end{center}

\begin{sloppypar}
	{\small \noindent
		M.~Abdullahi$^{ep,er}$              
		M.~Abrate$^{ek,el}$,                
		J.H.~Adams Jr.$^{ld}$,              
		D.~Allard$^{cb}$,                   
		P.~Alldredge$^{ld}$,                
		R.~Aloisio$^{ep,er}$,               
		R.~Ammendola$^{ei}$,                
		A.~Anastasio$^{ef}$,                
		L.~Anchordoqui$^{le}$,              
		V.~Andreoli$^{ek,el}$
% ...
	}
\end{sloppypar}
\vspace*{.3cm}

\begin{spacing}{1.05}
{
	\footnotesize
	\noindent
	%
% ...
$^{cb}$ Universit\'e de Paris, AstroParticule et Cosmologie (CNRS), Paris, France\\
% ...
$^{ef}$ Istituto Nazionale di Fisica Nucleare (INFN), Sezione di Napoli, Naples, Italy\\
% ...
$^{ei}$ Istituto Nazionale di Fisica Nucleare (INFN), Sezione di Roma Tor Vergata, Rome, Italy\\
% ...
$^{ek}$ Istituto Nazionale di Fisica Nucleare (INFN), Sezione di Torino, Turin, Italy\\
	$^{el}$ Universit\`a di Torino, Dipartimento di Fisica, Turin, Italy\\
% ...
$^{ep}$ Gran Sasso Science Institute (GSSI), L’Aquila, Italy\\
	$^{er}$ Istituto Nazionale di Fisica Nucleare (INFN), Laboratori Nazionali del Gran Sasso, Assergi, Italy\\
% ...
$^{ld}$ University of Alabama in Huntsville, Huntsville, AL, USA\\
	$^{le}$ City University of New York (CUNY), Lehman College, Bronx, NY, USA\\
% ...
}
\end{spacing}

%

\end{document}